\documentclass{article}
\usepackage{ieee_tns}
\usepackage{epsf}       

\begin{document}
\title{
Performance of a Prototype  CMS Hadron Barrel Calorimeter
in a Test Beam
\thanks{
Representing the CMS HCAL Group at the IEEE Conference,
Toronto, Nov. 1998 (Univ. of Rochester Preprint
UR-1544 ER/40685/923).
}
}

\author{
Arie Bodek\\
Department of Physics \& Astronomy\\
         University of Rochester \\
         Rochester, NY 14627, USA
}
\maketitle

\begin{abstract}
We report on the performance of a prototype
CMS Hadron Barrel Calorimeter (HCAL) module in a test
beam. The prototype
sampling calorimeter used copper
absorber plates and  scintillator tiles
with wavelength shifting fibers for readout.
Placing a lead tungstate crystal  
electromagnetic calorimeter in front of
HCAL affects the linearity and energy resolution
of the combined system  to hadrons. 
The data are used to optimize the
choice of total absorber depth, sampling frequency,
and longitudinal readout segmentation.

\end{abstract}

\section{Introduction}

We report on a variety of tests  performed in a test beam
with a prototype CMS Hadron Barrel
Calorimeter~\cite{tp,tdr}(HCAL) module
in 1995-96.
The prototype
sampling calorimeter is constructed
from copper absorber plates and scintillator tiles  
with wave length shifter (WLS) fibers for
readout~\cite{cdf_end_plug_upgrade_book}.
Data were taken both with and without a prototype lead
tungstate electromagnetic calorimeter
(ECAL) placed upstream of the HCAL module.

In this report, we discuss the performance of HCAL, including the 
effects of the lead tungstate crystals on the 
linearity and energy resolution of the response of the
combined ECAL+HCAL system 
to hadrons. This study is used in the  optimization of the HCAL design,
including the choice of total absorber depth, sampling frequency,
and longitudinal readout segmentation.

The CMS Central HCAL calorimeter will operate inside a 4 Tesla magnetic field.
A measurement of the effects of a magnetic field
on the response of HCAL to muons, electrons and pions is
presented elsewhere.~\cite{calor97_kunori}

\section{Experimental Setup}

The CMS combined calorimetric ECAL+HCAL system was tested in 1995
and 1996  at the H2 and H4 CERN beamlines. Data were taken
using beams of muons,
electrons and hadrons, ranging in momenta from 15 to 375 GeV/c.
Test beam prototypes of HCAL were based on the 
hanging file structure~\cite{kryshkin}. Copper
($\lambda_{INT}$(Cu)=15.06 cm) absorber plates 
varying in thickness from 2~cm to 10~cm were interspersed with
scintillator
tiles read out with WLS fibers. The total interaction depth 
of HCAL prototypes corresponds to 8.5 $\lambda_{INT}$ (H4, 1995 module)
and  10 $\lambda_{INT}$ (H2, 1996 module).
The transverse size of the calorimeter
is 64~cm$\times$64~cm.
Each scintillator layer is read out
independently by conventional photomultiplier tubes (PMT).
The light yield of each scintillator layer corresponds to  approximately to 2 
photoelectrons per minimum ionizing particle (2 PE/mip). 
The relative calibration of scintillator layers is established to
accuracy of $\approx$ 3\%
by equalizing the average response of each layer to muons.
The average pulse height deposited by muons in HCAL
corresponds to approximately 4 GeV (equivalent hadronic energy).

The prototype ECAL detector consists of a matrix
of 7$\times$7 PbWO$_{4}$ crystals, 23~cm long (25.8 X$_0$, 
1.1$ \lambda_{INT}$), each 2~cm$\times$2~cm in transverse size.
The calibration of the ECAL crystals is done using 50 GeV/c electrons
directed into center of each crystal. The electronic noise of the 
7$\times$7 crystal
matrix energy sum had a RMS width equivalent to 440 MeV (much better performance
is expected in future prototypes).
The absolute energy calibration of HCAL 
is set using 50 GeV/c pions interacting in HCAL only.
The absolute energy scale of ECAL is set using
50 GeV/c electrons.

\section{ The $e/h$ response ratio of HCAL}

The performance of a hadron calorimeter, i.e. hadron 
energy response linearity and energy resolution,
depends~\cite{wigmans} on the intrinsic
$e/h$, the  ratio of the response to the
electromagnetic and  hadronic shower components. The average 
fraction of the electromagnetic
component, F($\pi^0$) in pion induced showers increases as a function
of incident energy. For 
non-compensating calorimeters ( $e/h$ $\neq$1), this implies a non-linear
hadron energy response. The event-by-event fluctuations
in F($\pi^0$) contribute to variations in the reconstructed
shower energy and at high energies dominate the energy resolution for
hadrons.

We have measured the linearity of the energy response of HCAL to
pions which did not interact in ECAL. 
There is approximately a 9\% increase in the response 
between 15 and 375 GeV/c. 
In a separate test, the response of HCAL to electrons was measured by moving
the ECAL module out of the beamline. The response of the HCAL
to electrons is linear within 2\% and approximately 20\% higher
than the average response to pions.

\begin{figure}
\begin{center}
\epsfxsize=3in
\mbox{\epsffile{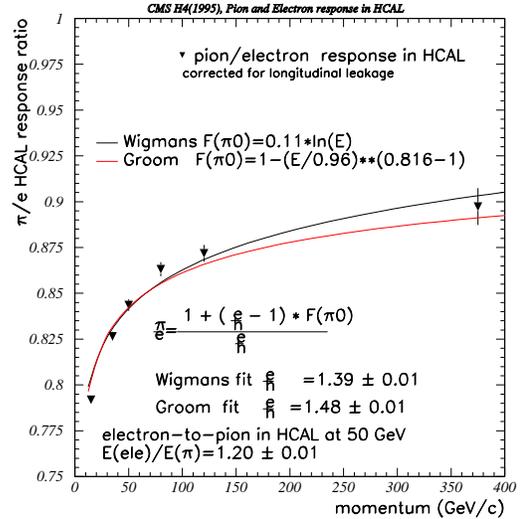}}
\caption{H4(1995) data: the pion/electron ratio of
response of the copper sampling prototype
HCAL as a function of beam momentum.
The calorimeter consists of the Inner HCAL (ten 3~cm Cu samplings followed 
by nine 6~cm Cu samplings),
the magnetic coil mimic (8~cm Cu+29~cm Al) and the HCAL Outer 
(two 8~cm Cu and two 10~cm Cu samplings). 
The scintillator is 4~mm thick SCSN-81.
The pion response of HCAL has been corrected for longitudinal
leakage. 
We have assumed a linear electron response of HCAL, 
E(ele)/E($\pi$ at 50 GeV)=1.20$\pm$0.01. 
The extracted values of $e/h$ correspond to the two different
parameterizations of the average fraction of $\pi^0$'s produced
in pion induced showers.
Similar extractions of $e/h$ for iron/scintillator  sampling
calorimeter was done by the CDF End Plug Upgrade group, 
see J.B. Liu et al., 
CDF Coll., CALOR97 Proceedings.
}
\label{eh-b}
\end{center}
\end{figure}

Using these two  data sets (the response of HCAL to electrons
and response to pions which interacting in  HCAL only)
we extract the value of the intrinsic $e/h$ of HCAL.
Two available parameterizations for F($\pi^0$),
one by Wigmans~\cite{wigmans} and the
other by Groom~\cite{groom}, yield values of $e/h$ 
of 1.38 $\pm$ 0.01 and 1.49 $\pm$ 0.01, respectively ( see Fig.~\ref{eh-b}).
The simple Wigmans formula for F($\pi^0$) increases with energy
($\approx$ log(E)) and becomes non-physical at very high energies.
The Groom's parameterization of  GCALOR Monte Carlo simulations
uses an exponential form.

\section{Linearity and Energy Resolution of the PbWO$_4$ ECAL+HCAL}

\begin{figure}
\begin{center}
\epsfxsize=3in
\mbox{\epsffile{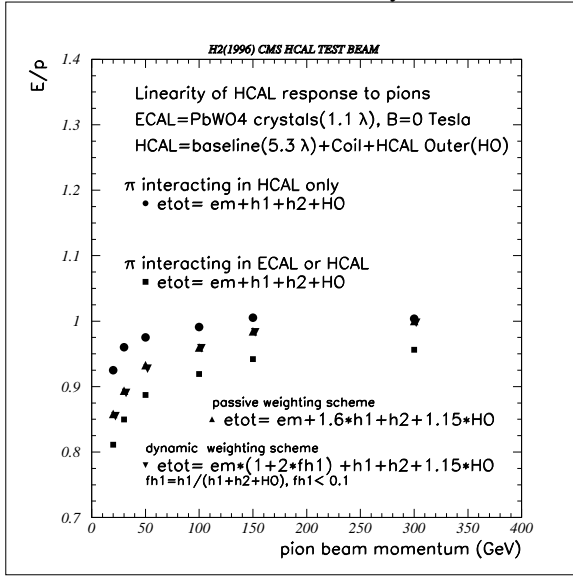}}
\caption{ The
linearity of the response to pions of
HCAL, and the combined PbWO$_4$ ECAL + HCAL system.
Statistical error bars are smaller than symbols.
The ECAL corresponds to 1.1$ \lambda_{INT}$ of lead tungstate crystals.
The HCAL readout corresponds to the Baseline Inner HCAL+HCAL Outer,
as defined below.
The Inner HCAL(5.3$ \lambda_{INT}$)
consists of two independent readouts: H1 (following
a 2~cm Cu plate) and H2 (thirteen 6~cm Cu samplings).
A single readout HCAL Outer (HO) consists of three samplings: 
immediately after the magnetic
coil (mimicked by 18~cm of Cu), followed by a 22~cm Cu sampling
and a 16~cm Cu sampling. The combined coil and HO samplings corresponds
to 2.5$ \lambda_{INT}$. The
scintillator plates are 4~mm thick SCSN-81.
}
\label{lin}
\end{center}
\end{figure}

\begin{figure}
\begin{center}
\epsfxsize=3in
\mbox{\epsffile{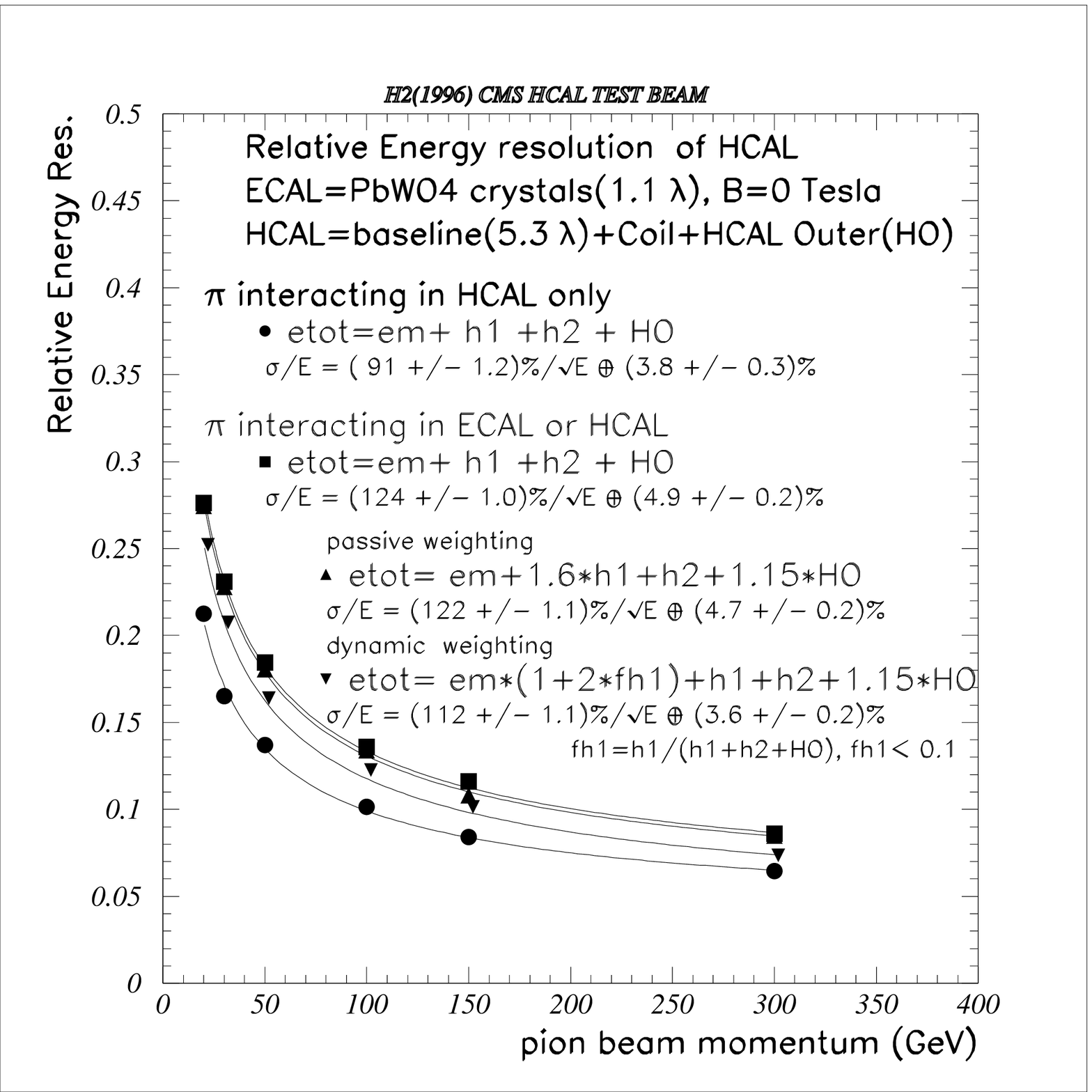}}
\caption{ The
relative pion energy resolution  of HCAL, and
the combined PbWO$_4$ ECAL + HCAL system. Statistical error bars
are smaller than symbols.
The ECAL corresponds to 1.1$ \lambda_{INT}$ of lead tungstate crystals.
The HCAL readout corresponds to the Baseline Inner HCAL+HCAL Outer,
as defined below.
The Inner HCAL(5.3$ \lambda_{INT}$) 
consists of two independent readouts: H1 (following
a 2~cm Cu plate) and H2 (thirteen 6~cm Cu samplings).
A single readout HCAL Outer (HO) consists of three samplings: 
immediately after the magnetic
coil (mimicked by 18~cm of Cu), followed by a 22~cm Cu sampling
and a 16~cm Cu sampling. The combined coil and HO samplings corresponds
to 2.5$ \lambda_{INT}$. The scintillator plates are 4~mm thick SCSN-81.
}
\label{res}
\end{center}
\end{figure}

Figure~\ref{lin} and~\ref{res} show comparisons of
the energy response and relative energy resolutions
of the combined PbWO$_4$ ECAL+HCAL calorimeters to pions. 
The inclusive set of  pions which
interact in either ECAL or HCAL have approximately 10\% lower
response than the response of those pions
which interact in HCAL only.
In addition, the
relative energy resolution of the combined calorimetric system
is significantly degraded, as compared to the energy resolution
of the standalone HCAL system. We attribute these results
to the large $e/h$ of the PbWO$_4$ ECAL system.

   We have tested two possible approaches to correct for the degradation
   of the performance of the combined ECAL+HCAL calorimeters.
   Both of these methods make use of the segmented readout of the 
   HCAL. The first readout segment (H1) covers a sampling
   immediately downstream of the ECAL. The second readout segment (H2) 
   includes the remainder of HCAL inside the magnet coil.
    A third readout segment
   (HO) corresponds to the HCAL sampling located outside of the 
   magnet cryostat.
 
   In the first approach, called
   passive weighting, we reduce the
   non-linearity of energy response (E/p) and  relative energy resolution 
   (rms(E)/E) by increasing the weight ($\alpha$) of the first (H1) HCAL 
   readout segment, where $\alpha$ is energy independent constant.

\begin{equation}
    E_{TOT} = E_{ECAL} + \alpha \times E_{H1} + E_{H2} + E_{HO}
\end{equation}

   In the second approach  we use a dynamic
   correction to reduce the effect of the 
   large $e/h$ value of the PbWO$_{4}$ ECAL calorimeter.
   This event by event correction, which depends
    on the fraction of the energy
   deposited in the first readout segment of HCAL immediately 
   downstream of the ECAL
   effectively allows one to have a larger
   correction for the low ECAL response to pions
   interacting in ECAL. 

   \begin{equation}
    E_{TOT} = (1+2\times f(H1))\times E_{ECAL} + E_{H1} + E_{H2} + E_{HO},
   \end{equation}
   \begin{equation}
   f(H1)=E(H1)/(E(H1)+E(H2)+E(HO)),~f(H1) \leq 0.1
   \end{equation}

   Using either the passive or dynamic correction technique, 
   one finds  a residual energy nonlinearity
   of only 10\% for pions with energy between 30 and 300 GeV/c.
   Note that while the passive weighting method
   can be applied to single particles as well as to the multiparticle
   events (jets), the dynamic weighting method may result in
   high energy tails for $\pi^0$$\pi^{\pm}$ multiparticle jets. If one
   uses the passive weighting scheme, then
   a separate readout for H1 is not needed.

\section{Optimization of the HCAL design}

During the H2(1996) tests, the prototype HCAL module was segmented
longitudinally into 27 readout layers.
Using the test beam data, we have simulated various sampling configurations
in order to study the
performance of HCAL as a function of the
total interaction length and sampling frequency.
The average longitudinal profiles of pions extend past the
magnetic coil. This motivated the decision to instrument the CMS muon system 
iron absorbers (located outside the coil) with scintillator plates
for use  as a pion shower tail catchers. We refer to
this part of the calorimeter as the HCAL Outer (HO) system.
By adding the HO, the gaussian parts, as well as the non-gaussian
low energy tails in the energy distributions are significantly reduced.

We have also compared relative energy resolution of the combined
PbWO$_4$ ECAL+HCAL system for three cases of the HCAL absorber samplings:
3~cm Cu sampling for the first eight layers followed by 6~cm Cu
sampling, 6~cm Cu uniform sampling and 12~cm Cu uniform sampling.
The study indicates that the energy resolution is not dominated
by the sampling fluctuations in HCAL, but by the poor $e/h$ of the
PbWO$_{4}$ ECAL system.
Therefore a factor of two change in the sampling
frequency ( 3~cm/6~cm vs 6~cm uniform sampling)
of HCAL does not result in a noticeable
degradation of the energy resolution of the detector. In the case 
of a 12~cm Cu sampling, the degradation in energy resolution is
noticeable, but did  not scale with $\sqrt{t}$ ( where $t$ is the
thickness of the absorber plates). This indicates that even in  high 
pseudorapidity regions
of the calorimeter, the sampling fluctuations do not dominate
the overall resolution of the ECAL+HCAL system.

\end{document}